# In$_{1-x}$Mn$_x$Sb - a new narrow gap ferromagnetic semiconductor

T. Wojtowicz[a], and G. Cywiński

Department of Physics, University of Notre Dame, Notre Dame, IN 46556 and
Institute of Physics, Polish Academy of Sciences, 02-668 Warsaw, Poland

W.L. Lim, X. Liu, M. Dobrowolska, and J. K. Furdyna

Department of Physics, University of Notre Dame, Notre Dame, IN 46556

K. M. Yu and W. Walukiewicz

Electronic Materials Program, Materials Sciences Division,
Lawrence Berkeley National Laboratory, Berkeley, CA 94720

G.B. Kim, M. Cheon, X. Chen, S.M. Wang, and H. Luo

Department of Physics University at Buffalo, State University of New York, Buffalo, NY 14260

ABSTRACT

A narrow-gap ferromagnetic In$_{1-x}$Mn$_x$Sb semiconductor alloy was successfully grown by low-temperature molecular beam epitaxy on CdTe/GaAs hybrid substrates. Ferromagnetic order in In$_{1-x}$Mn$_x$Sb was unambiguously established by the observation of clear hysteresis loops both in direct magnetization measurements and in the anomalous Hall effect, with Curie temperatures T$_C$ ranging up to 8.5 K. The observed values of T$_C$ agree well with the existing models of carrier-induced ferromagnetism.

---

[a] Author to whom correspondence should be addressed; electronic mail: wojto@ifpan.edu.pl

The intense research activity in the area of ferromagnetic semiconductor alloys [1,2] has largely focused on III-Mn-V alloys with small lattice constants and large effective masses of valence-band holes, including the most thoroughly explored $Ga_{1-x}Mn_xAs$, and more recently $Ga_{1-x}Mn_xN$ and $In_{1-x}Mn_xN$ [3]. This can probably be ascribed to the expectation that materials with the above parameters are likely to yield the highest values of the Curie temperature $T_C$ [2]. Bearing in mind, however, that actual mechanism of ferromagnetism in III-Mn-Vs is still not entirely clear [3,4], it is important to explore the opposite extreme of the III-Mn-V ternaries: i.e., $In_{1-x}Mn_xSb$, which has the largest lattice constant and the smallest hole effective mass in this family of materials.

In addition to shedding light on the magnetic trends associated with lattice and band parameters, the smallness of the energy gap of InSb-based alloys offers the possibility of large *relative* magnetic-field tuning of the bandgap, with possible applications in infrared spin-photonics. Moreover, due to the smallness of the hole effective masses, InSb-based alloys hold promise of high hole mobilities, of interest in applications involving carrier transport. To date, only the quaternary $In_{1-x}Mn_xAs_{1-y}Sb_y$ ferromagnetic system has been produced by metal organic MBE, and interesting light-induced changes of magnetization have been reported for this multi-component alloy[5].

In this paper we report the growth and characterization of $In_{1-x}Mn_xSb$ ferromagnetic semiconductor in thin layer form. The physical properties of $In_{1-x}Mn_xSb$ layers were studied by x-ray diffraction, SQUID magnetometry, electrical transport, and Rutherford back scattering and particle-induced X-ray emission in the channeling configuration (the latter to determine the Mn site location in the lattice) [6].

The In$_{1-x}$Mn$_x$Sb films were grown in a Riber 32 R&D molecular beam epitaxy (MBE) system at Notre Dame. Fluxes of In and Mn were supplied from standard effusion cells, and Sb$_2$ flux was produced by an Sb cracker cell. To avoid large parallel conductance that occurs in thick InSb buffer grown at high temperature, the growth was performed on closely lattice matched (001) hybrid CdTe/GaAs substrates grown by MBE at the Institute of Physics of the Polish Academy of Sciences. Prior to film deposition we grew a 100 nm low temperature (LT) InSb buffer layer at 210 ºC, which provided a flat substrate surface for subsequent deposition, as evidenced by a well resolved streaky RHEED pattern. The substrate was then cooled to 170 ºC for the growth of a 230-nm-thick LT-In$_{1-x}$Mn$_x$Sb or LT- In$_{1-y}$Be$_y$Sb. The Sb$_2$:In beam equivalent pressure ratio of 3:1 was used for both nonmagnetic and magnetic materials. During the growth of LT-In$_{1-x}$Mn$_x$Sb and LT- In$_{1-y}$Be$_y$Sb we observed a streaky RHEED pattern, indicative of two dimensional growth with 1×3 reconstructions. The growth rate (0.26 ml/s), and hence the thickness of the layers, was determined from intensity oscillations of the RHEED specular spot.

The crystallographic quality of the In$_{1-x}$Mn$_x$Sb/InSb/CdTe layers as well as the variation of the In$_{1-x}$Mn$_x$Sb lattice constant with x were determined by x-ray diffraction. Figure1 shows θ-2θ x-ray diffraction curves for the (004) reflection obtained using Cu-*Kα* radiation for a series of structures. The shift of the diffraction peaks of In$_{1-x}$Mn$_x$Sb to larger angles for increasing x indicates that the lattice constant of the magnetic layer decreases with x. Quantitatively the relaxed lattice constant decreases from *d$_0$* = 0.64794 nm in InSb to 0.64764 nm and 0.64762 nm in layers with x=0.02 and 0.028, respectively. The full width at half maximum (FWHM) of the In$_{0.972}$Mn$_{0.028}$Sb rocking curve (see inset to Fig. 1) was 206 arc-sec (as compared to 160 arc-sec for the LT-InSb/CdTe peak), indicating good crystal quality of the magnetic alloy. No sign of other crystallographic phases (e.g., MnSb) was detected in x-ray studies for the range of compositions discussed here.

The results of DC magnetotransport experiments performed in a six-probe Hall bar geometry on samples with ohmic In contacts are shown in Fig. 2. Temperature dependence of zero-field resistivity $\rho$ for ferromagnetic $In_{1-x}Mn_xSb$ with various nominal values of x (as determined by the temperature of the Mn effusion cell) is presented in Fig. 2a. The data for one non-magnetic $In_{1-y}Be_ySb$ layer with a similar hole concentration $p$ is also given for comparison (the value of $p$ in $In_{1-y}Be_ySb$ was determined by Hall measurement to be $1.4 \times 10^{20}$ cm$^{-3}$). One can see that while the resistivity of the nonmagnetic sample decreases monotonically with decreasing temperature (characteristic of metallic conductivity) due to increasing hole mobility, the $\rho$ vs. $T$ dependence for all magnetic layers is nonmonotonic. Note in particular that the resistivities of the three $In_{1-x}Mn_xSb$ layers grown with Mn cell temperatures $T_{Mn}$ of 700, 710 and 720 ºC show clear maxima at $T_\rho$ of 5.4, 7.5 and 9.2 K, respectively. These maxima signal the paramagnetic-to-ferromagnetic phase transition. It is known from studies of $Ga_{1-x}Mn_xAs$ that $T_\rho$ occurs close to the Curie temperature [1], thus serving as a good estimate of $T_C$ [7]. On theoretical grounds the occurrence of the maximum in $\rho(T)$ can be understood in terms of a model that combines scattering by magnetoimpurities and magnetic fluctuations [8,9]. To our knowledge, such clear maxima were not yet observed in any III-Mn-V ferromagnetic semiconductor other than $Ga_{1-x}Mn_xAs$, and therefore studies of electrical transport in $In_{1-x}Mn_xSb$ can serve as an important test case for theories aiming at a quantitative description of critical scattering in this class of ferromagnetic materials.

The appearance of ferromagnetic state in mixed $In_{1-x}Mn_xSb$ crystal at temperatures near (i.e., slightly below) $T_\rho$ was unambiguously confirmed by our studies of the Hall effect.

It is well known that in ferromagnetic materials the Hall resistivity can be expressed as a sum of two contributions: $\rho_{Hall} = R_0 B + R_A M$, where $R_0$ and $R_A$ are the ordinary and the anomalous Hall coefficients, respectively, $B$ is the applied magnetic field, and $M$ is the magnetization [1,10]. The observation of the anomalous Hall effect (AHE) and, even more importantly, of clear hysteresis loops in AHE, is decisive for establishing ferromagnetism in any new material. This is because the presence of ferromagnetic characteristics in SQUID measurements, as opposed to those in electrical transport, could also be caused by ferromagnetic precipitates (e.g., MnSb) in addition to (or instead of) a homogeneous ferromagnetic alloy. Figure 2b shows the Hall resistivity $\rho_{Hall}$ vs. $B$ measured in four $In_{1-x}Mn_xSb/InSb/CdTe$ layers at 1.4 K. Here one should note that both the amplitude of the AHE and the width of the hysteresis loops increase with increasing Mn concentration. The sign of the AHE coefficient $R_A$ is found to be negative (opposite to that of the normal Hall coefficient of p-type nonmagnetic layers, e.g., our $In_{1-y}Be_ySb$) for all $In_{1-x}Mn_xSb$ samples studied, independent on Mn and/or hole concentration.

Magnetization of $In_{1-x}Mn_xSb$ obtained via SQUID measurements is shown in Fig. 3. Here we plot $M$ vs. $B$ for several temperatures, observed with $B$ applied either perpendicular or parallel to the layer plane. As in the AHE data, clear hysteresis loops are observed. One can also see in the data for T = 2 K that $M$ saturates at higher fields when $B$ is applied in the layer plane, indicating that -- as expected for the small tensile strain in the magnetic layer -- the easy axis of magnetization is perpendicular to the layer plane. The inset in Fig. 3 shows the temperature dependence of the remanent magnetization $M$ measured in a perpendicular field of 10 G in two $In_{1-x}Mn_xSb$ layers (x = 0.02 and 0.028). The Curie temperatures determined from these plots, $T_C = 7 \pm 0.5$ K and $8.5 \pm 0.5$ K, are in good agreement with $T_\rho = 7.5$ K and 9.2 K, respectively, for the two samples.

Having unambiguously established that $In_{1-x}Mn_xSb$ belongs to the class of diluted ferromagnetic semiconductors, it would be interesting to learn whether currently available models of carrier mediated ferromagnetism correctly describe also this narrowest-energy-gap member of the III-Mn-V family of materials. To our knowledge, there is only one paper that considers theoretically ferromagnetic properties of $In_{1-x}Mn_xSb$ and predicts values of its Curie temperatures [11,12]. Starting from the simplest mean-field theory, this model estimates the $T_C$ enhancement due to exchange and correlation in the itinerant-hole system, and the $T_C$ suppression due to collective fluctuations of the ordered moment. For a given host material two parameters are crucial for the model: concentration of $Mn^{++}$ 5/2 spins that participate in the ferromagnetism; and the concentration of free holes that mediate this interaction through the *p-d* coupling. It was theoretically shown that the *p-d* coupling between the holes and the Mn ions at interstitial sites ($Mn_I$) is negligible [13], so that $Mn_I$ do not participate in the hole-mediated ferromagnetism [7]. Additionally, because of strong antiferromagnetic coupling between substitutional Mn at the In sites ($Mn_{In}$) and $Mn_I$ [13] which are brought together by electrostatic interaction to form $Mn_I$-$Mn_{In}$ pairs [6], the concentration of ferromagnetically active $Mn_{In}$, $x_{eff}$, is further reduced to the value $x_{eff} = (x_{In} - x_I)x$, where $x_{In}$ and $x_I$ are, respectively, substitutional and interstitial atomic fractions of Mn. Theoretical calculations for $x = 0.028$ with $x_{eff} = 0.021$ determined from our c-PIXE/RBS [6] experiments and with experimentally estimated free hole concentration $p_{Hall} = 2.1 \times 10^{20}$ cm$^{-3}$ give a Curie temperatures $T_C$ of 8 K, in good agreement with the experimentally determined value of $8.5 \pm 0.5$ K. The free hole concentration used above was estimated from the slope of the Hall resistivity measured at 1.4 K in magnetic fields above 5.5 T. It was found that under these conditions magnetization is saturated (within our experimental error) and the magnetoresistance is negligible, thus resulting in only a very small AHE contribution.

In summary, we have successfully grown a new narrow-gap ferromagnetic semiconductor alloy $In_{1-x}Mn_xSb$ – a material which possesses all attributes of carrier-mediated ferromagnetism, including a well-defined peak in the temperature dependence of resistivity, and clear hysteresis loops observed in both AHE and in direct SQUID magnetization. Since $In_{1-x}Mn_xSb$ is characterized by the largest lattice constant, and by band parameters which are quite different from those of other III-V-based ferromagnetic semiconductors, the magnetic properties of $In_{1-x}Mn_xSb$ can be used for testing current theoretical models of ferromagnetism in III-Mn-V ferromagnets generally. The material also holds promise of those spintronic applications where either small masses or high carrier mobilities may be desirable.

This work was supported by the DARPA SpinS Program under ONR grant N00014-00-1-0951; by the Center of Excellence CELDIS established under EU Contract No. ICA1-CT-2000-70018 (Poland), by the Director, Office of Science, Office of Basic Energy Sciences, Division of Materials Sciences and Engineering, of the U.S. Department of Energy under Contract No. DE-AC03-76SF00098; and by NSF Grant DMR00-72897.

FIGURE CAPTIONS

Fig. 1  θ-2θ x-ray diffraction scans for $In_{1-x}Mn_xSb/InSb/CdTe$ layers with x = 0, 0.02, 0.028 obtained using Cu $K\alpha$ radiation for the (004) reflection. The inset shows the rocking curve for $In_{0.972}Mn_{0.028}Sb/InSb/CdTe$.

Fig. 2  a) Temperature dependence of zero-field resistivity ρ for $In_{1-x}Mn_xSb/InSb/CdTe$ ferromagnetic layers with various x as determined by the temperature of the Mn effusion cell during the growth (indicated in the figure by $T_{Mn}$). The data for one non-magnetic $In_{1-y}Be_ySb/InSb/CdTe$ layer with a similar hole concentration is also given for comparison (open circles). In order to avoid elongating the scale of the figure the ρ of the non-magnetic layer is shifted up by 0.1 mΩcm. b) Hall resistivity $ρ_{Hall}$ vs. magnetic field at 1.4 K measured in four $In_{1-x}Mn_xSb/InSb/CdTe$ layers. The data for the different samples were shifted along the y-axis; their order and symbols used are the same as in the left panel.

Fig. 3  Field dependence of magnetization measured by SQUID in $In_{0.972}Mn_{0.028}Sb$ at various temperatures with the field applied either perpendicular or parallel to the layer plane. The inset shows the temperature dependence of magnetization measured with a perpendicular field of 10 Gs for $In_{1-x}Mn_xSb$ with x = 0.02 and 0.028.

**REFERENCES**

[1] H. Ohno, Science **281**, 951 (1998).

[2] T. Dietl, H. Ohno, F. Matsukura, J. Cibert, and D. Ferrand, Science **287**, 1019 (2000).

[3] T. Dietl, Semicond. Sci. Technol. **17**, 377 (2002), and other papers in this special issue.

[4] K. Sato and H. Katayama-Yoshida, Semicond. Sci. Technol. 367 (2002).

[5] Y. K. Zhou, H. Asahi, S. Okumura, M. Kanamura, J. Asakura, K. Asami, M. Nakajima, H. Harima, and S. Gonda, J. Crystal Growth **227-228**, 614 (2001).

[6] K. M. Yu, W. Walukiewicz, T. Wojtowicz, I. Kuryliszyn, X. Liu, Y. Sasaki, and J. K. Furdyna, Phys. Rev. B **65**, 201303 (2002).

[7] T. Wojtowicz, W. L. Lim, X. Liu, Y. Sasaki, U. Bindley, M. Dobrowolska, J. K. Furdyna, K. M. Yu, and W. Walukiewicz, J. of Superconductivity: Incorporating Novel Magnetism (2002)- in press.

[8] E. L. Nagaev, Phys. Reports **346**, 387 (2001).

[9] Sh. U. Yuldashev, Hyunsik Im, V. Sh. Yalishev, C. S. Park, T. W. Kang, S. Lee, Y. Sasaki, X. Liu, and J. K. Furdyna, Appl. Phys. Lett. (2003)- in press.

[10] T. Jungwirth, Qian Niu, and A. H. MacDonald, Phys. Rev. Lett. **88**, 207208-1 (2002).

[11] T. Jungwirth, J. Konig, J. Sinova, J. Kucera, and A. H. MacDonald, Phys. Rev. B **66**, 012402/1 (2002).

[12] Ferromagnetic Semiconductor Spintronics Web Project, http://unix12.fzu.cz/ms (2003).

[13] J. Blinowski and P. Kacman, Phys. Rev. B (2003)- in press.

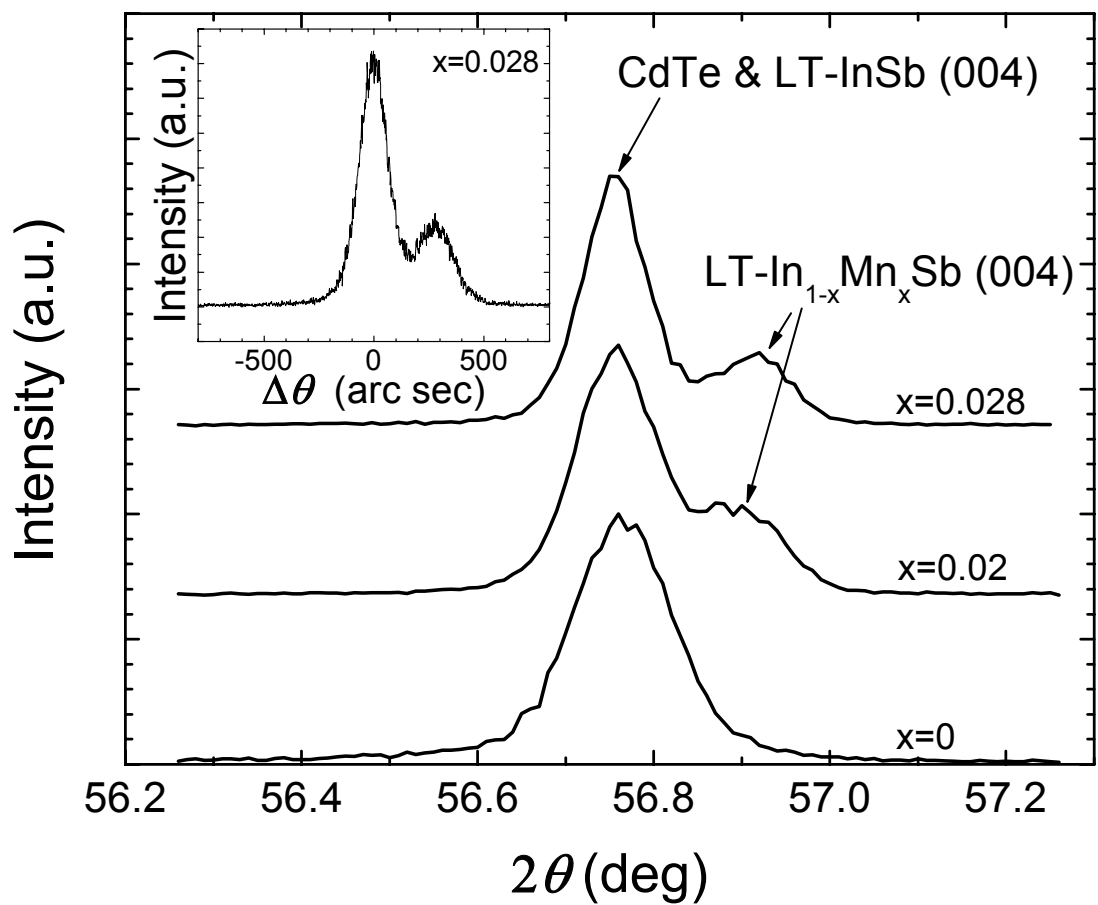

Fig. 1

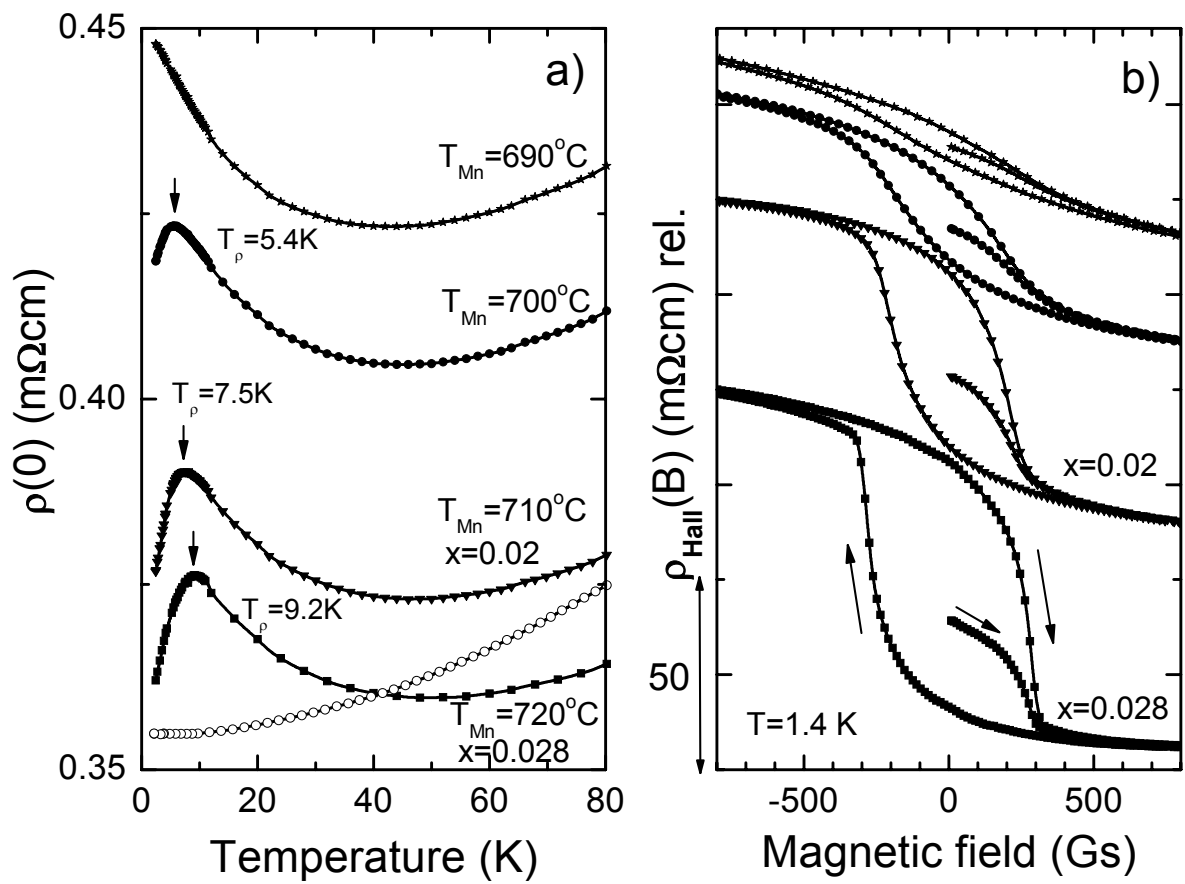

Fig. 2

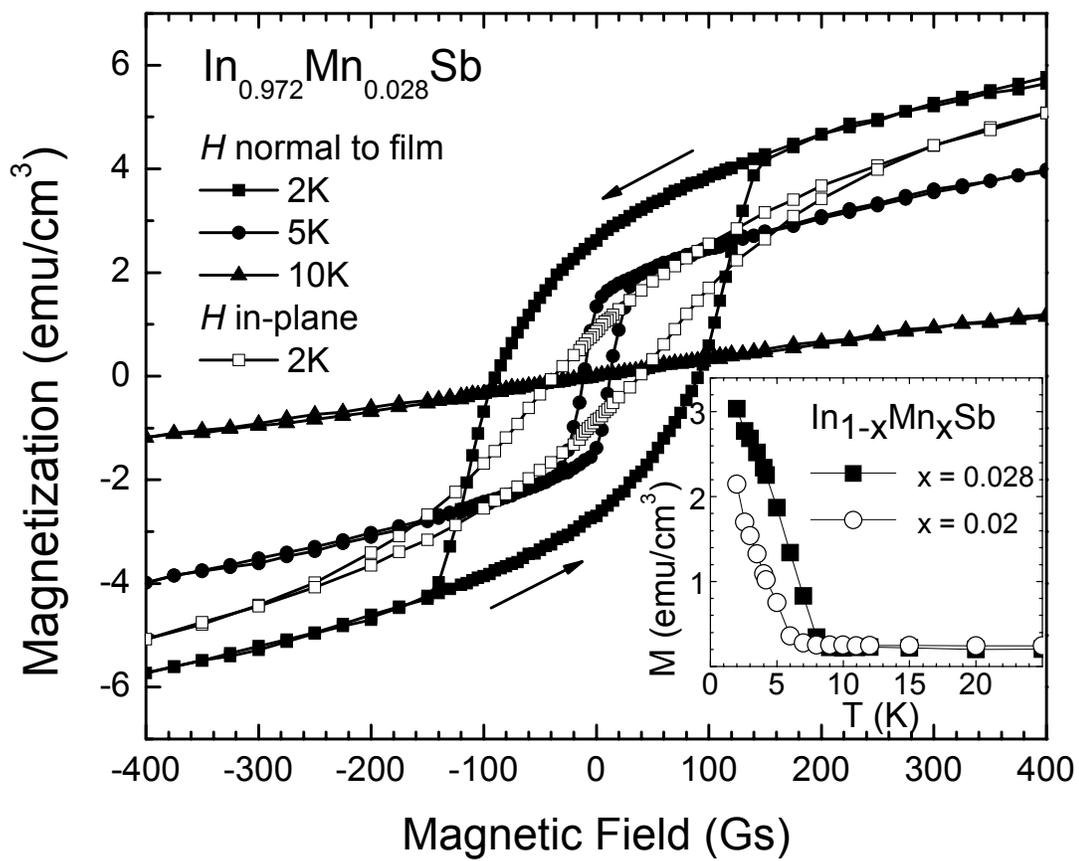

Fig. 3